\documentclass[a4paper,11pt]{article}
\usepackage{jcappub}
\usepackage{float} 

\usepackage{xspace}
\newcommand{\CNB}{C$\nu$B\xspace}
\title{Impact of Warm Dark Matter on the Cosmic Neutrino Background Anisotropies}

\author[1]{Christopher G.\ Tully}
\author[2]{and Gemma Zhang}
\affiliation[1]{Department of Physics, Princeton University, Princeton, New Jersey 08544, USA}
\affiliation[2]{Department of Physics, Harvard University, Cambridge, Massachusetts 02138, USA}
\emailAdd{cgtully@princeton.edu}
\emailAdd{yzhang7@g.harvard.edu}

\abstract{
The Cosmic Neutrino Background (\CNB) anisotropies for massive neutrinos are a unique probe of large-scale structure formation.  The redshift-distance measure is completely different for massive neutrinos as compared to electromagnetic radiation.  The \CNB anisotropies in massive neutrinos grow in response to non-relativistic motion in gravitational potentials seeded by relatively high $k$-modes.  Differences in the early phases of large-scale structure formation in Warm Dark Matter (WDM) versus Cold Dark Matter (CDM) cosmologies have an impact on the magnitude of the \CNB anisotropies for contributions to the angular power spectrum that peak at high $k$-modes.  We take the examples of WDM consisting of 2, 3 or 7\,keV sterile neutrinos and show that the \CNB anisotropies for 0.05\,eV neutrinos drop off at high-$l$ multipole moment in the angular power spectrum relative to CDM.  
At the same angular scales that one can observe baryonic acoustical oscillations in the CMB, the \CNB anisotropies begin to become sensitive to differences in WDM and CDM cosmologies.
The precision measurement of high-$l$ multipoles in the \CNB neutrino sky map is a potential possibility for the PTOLEMY experiment with thin film targets of spin-polarized atomic tritium superfluid that exhibit significant quantum liquid amplification for non-relativistic relic neutrino capture.
}

\makeatletter
\gdef\@fpheader{}
\makeatother
\begin{document}

\maketitle
\flushbottom


\section{Introduction}

The ``null assumption'' of the cosmological principle is that the universe on large scales is homogeneous and isotropic.  Deviations from this assumption are therefore a measure of large-scale structure formation.  Perhaps one of the most salient anisotropies appears in the Cosmic Microwave Background (CMB) and through a detailed modeling of the early universe tells us the fraction of baryonic matter to the total matter in the universe.  The baryonic matter is distinguished from the total matter through the formation of sound waves in a relativistic plasma before the period of plasma recombination~\cite{kamionkowski1999cosmic}.  The nature of the non-plasma/dark matter is largely unknown with an important exception being the fraction of neutrinos which are predicted from the Cosmic Neutrino Background (\CNB).  Anisotropies in the \CNB massive neutrinos are also an important measure of large-scale structure formation and reveal an earlier period of dynamics than viewed at equivalent distances using optical probes.

As described in Ref.~\cite{2021JCAP...06..053T}, the \CNB anisotropies for massive neutrinos grow rapidly during the non-relativistic phase of their motion.  With directionality, as proposed by the PTOLEMY experiment with a polarized tritium target, the \CNB neutrino capture rate variation provides a map of the anisotropies on the neutrino sky as seen on Earth.  The variations on the neutrino sky are therefore a measure of the total matter power spectrum at redshifts relevant to \CNB anisotropy growth.  Depending on the model for dark matter, the total matter power spectrum can differ during the period of \CNB anisotropy growth.  In particular, warm dark matter consisting of 2--7\,keV sterile neutrinos has a different contribution to the \CNB anisotropies as compared to cold dark matter.  Cold dark matter is assumed to have exactly zero pressure at high redshift while warm dark matter carries some degree of non-zero radiation pressure according to its mass~\cite{de2017equation,hipolito2018general}.

\section{\CNB Angular Power Spectra}

\begin{figure}[H]
\begin{center}
\includegraphics[width=0.65\linewidth]{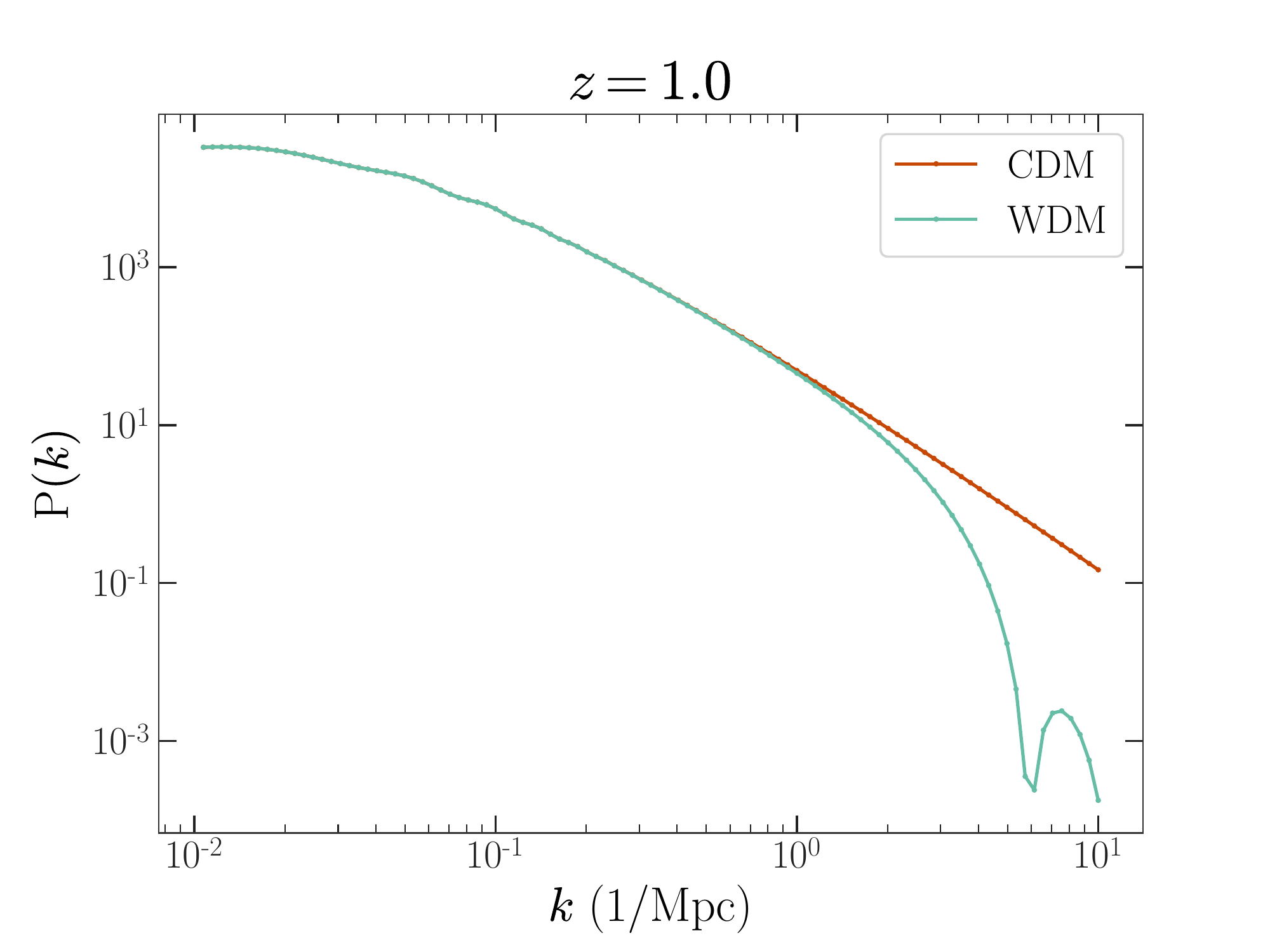}
\caption{The total matter linear power spectra for CDM and WDM with a 2\,keV sterile neutrino.\label{fig:Pk}}
\end{center}
\end{figure}

\begin{figure}[H]
\begin{center}
\includegraphics[width=0.65\linewidth]{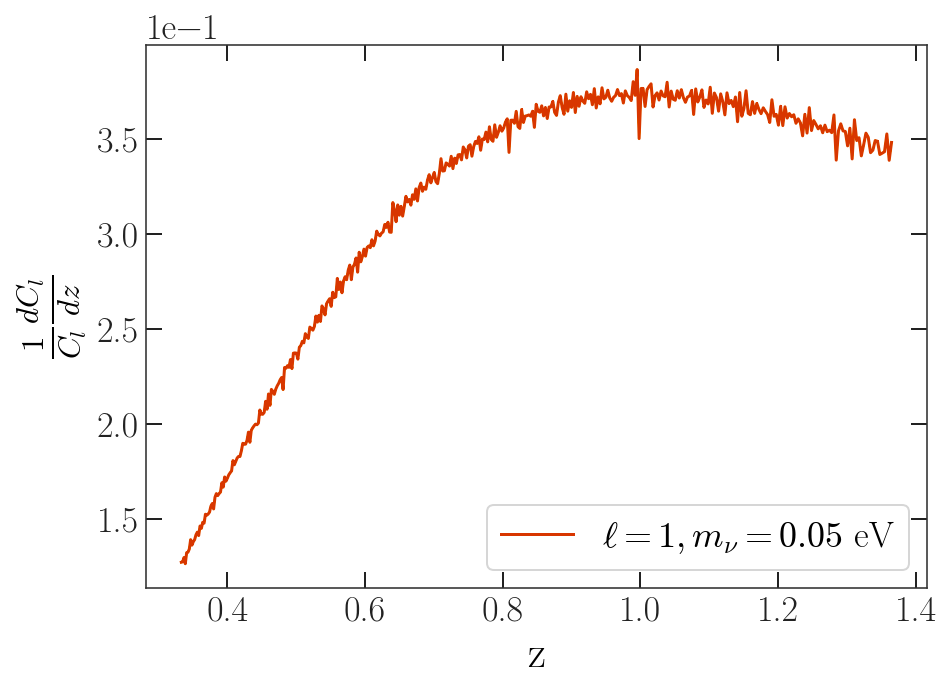}
\caption{The fractional contribution to the $l=1$ multipole of the \CNB angular power spectrum for a neutrino mass of 0.05\,eV as a function of redshift.\label{fig:Clz}}
\end{center}
\end{figure}

\begin{figure}[H]
\begin{center}
\includegraphics[width=0.49\linewidth]{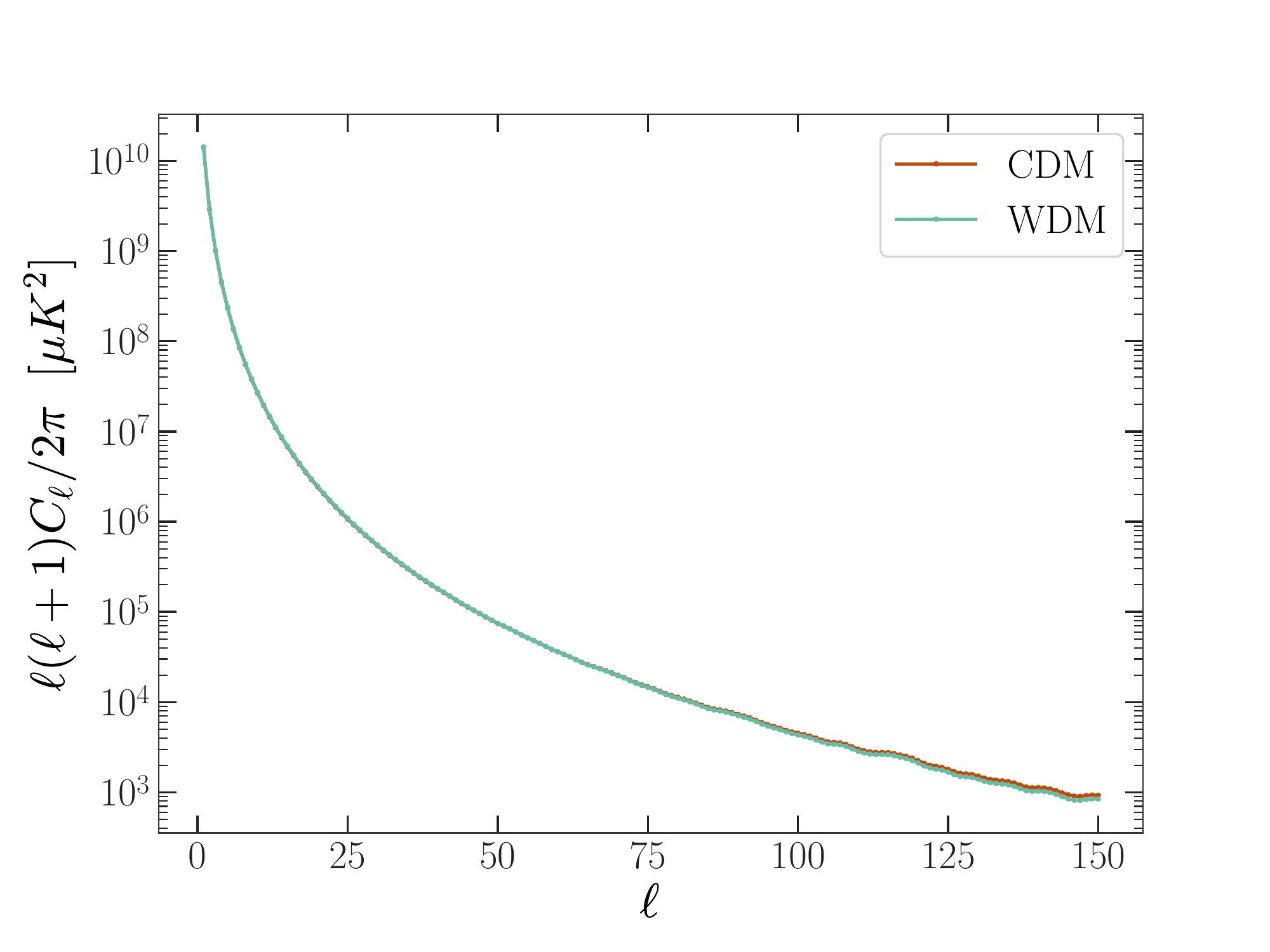}
\includegraphics[width=0.49\linewidth]{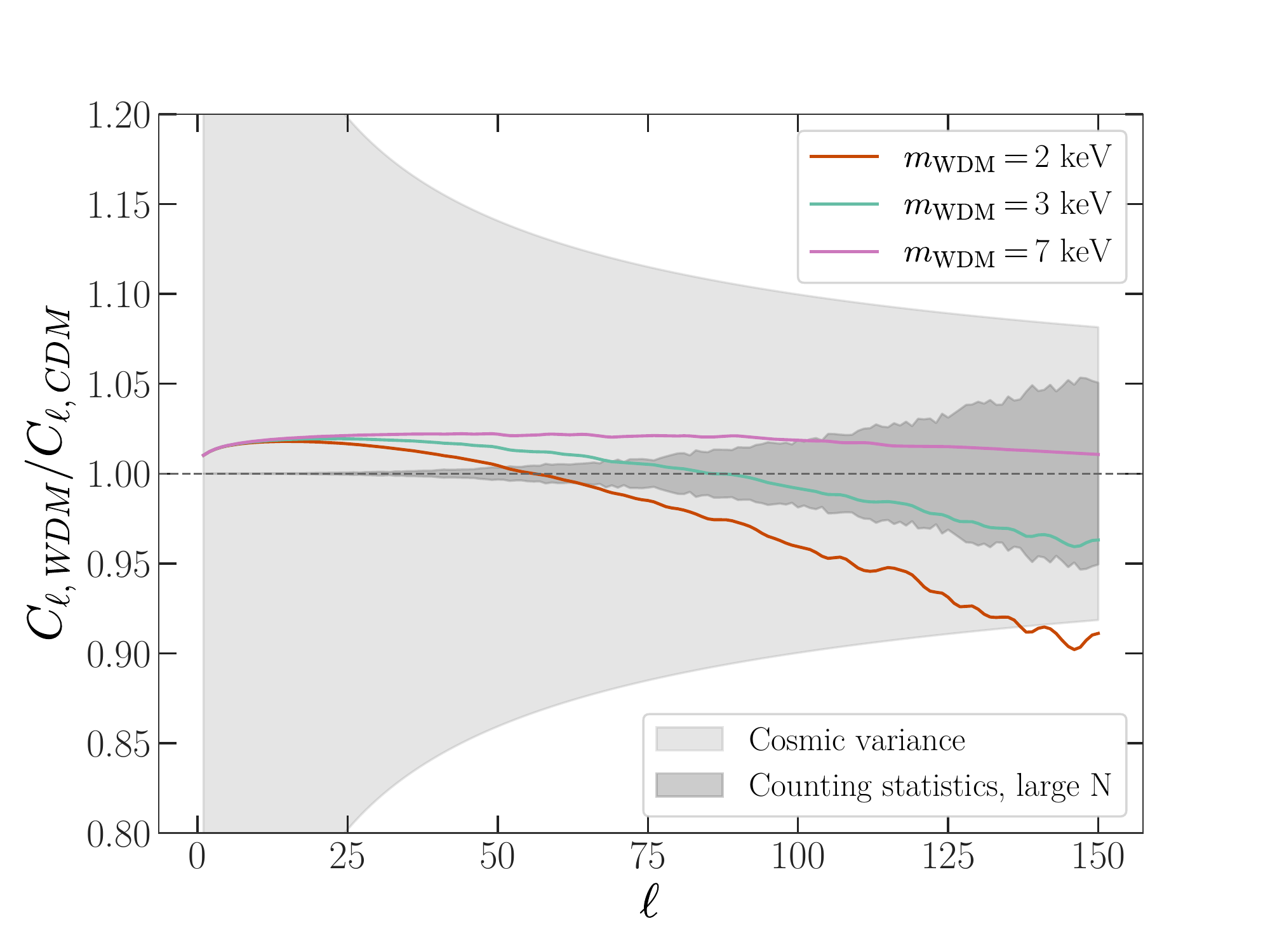}
\caption{{(\it left)} The total \CNB angular power spectra for CDM and WDM with a 2\,keV sterile neutrino. {\it (right)} The ratio of the \CNB angular power spectra for CDM and WDM cosmologies for a sterile neutrino mass of 2, 3 or 7\,keV.  The impact of WDM is a relative drop in the \CNB angular power at high-$l$.  The contributions to the total uncertainty on the measurement of the ratio are shown for cosmic variance and the PTOLEMY counting statistics in the limit of large-$N$ total neutrino capture events.\label{fig:Clratio}}
\end{center}
\end{figure}

The total matter linear power spectra are shown in Figure~\ref{fig:Pk} with the WDM modeled as a 2\,keV sterile neutrino in additional to a normal mass hierarchy of active neutrinos, the lightest massless, and computed using the CLASS software implemented within CO{\it N}CEPT~\cite{class2011,dakin2019}.  The CDM is modeled with the same normal mass hierarchy of active neutrinos.
At a redshift of $z=1$, the warm and cold dark matter models predict a different amount of linear power for $k$-modes of approximately 1\,Mpc$^{-1}$ and higher.
The differences in WDM and CDM in the high $k$-modes become less distinct at low redshift under the effects of non-linear dynamics~\cite{viel2012non}.
The redshifts relevant to the \CNB anisotropies for neutrinos with a mass of 0.05\,eV are shown in Figure~\ref{fig:Clz}.  The bulk of the fractional contribution to the total \CNB angular power spectra for a neutrino mass of 0.05\,eV come from redshifts of $z=1$ and higher.
The ratio of the \CNB angular power spectra for CDM and WDM with a 2, 3 or 7\,keV sterile neutrino is shown in Figure~\ref{fig:Clratio}.  Differences can be seen in the high-$l$ modes.

\unskip

\unskip

\unskip

The total uncertainty on the measurement of the \CNB angular power spectrum has contributions from cosmic variance and the PTOLEMY counting statistics for the total number of neutrino capture events.  The fractional one-standard-deviation uncertainty on $C_\ell$ from cosmic variance is given by
\begin{equation}
\frac{ \Delta C_\ell}{C_\ell} = \sqrt{\frac{2}{2l+1}} \qquad \mbox{(cosmic variance)}
\end{equation}
where for $l=150$, the fractional uncertainty is 8.2\%~\cite{peebles1980large}.  The fractional uncertainty on the measurement of $C_\ell$ from the PTOLEMY counting statistics per $l$-mode was determined numerically.  The procedure was to generate a temperature map, as in Ref.~\cite{2021JCAP...06..053T}, and to interpret the map as a count rate map for neutrino captures, without folding in the angular smearing from the polarized tritium target.  The expected count per pixel on the sky is then smeared using Poisson statistics.  The map is then inverted to find the $C_\ell$ and this procedure is repeated for several trials.  We verified that the fractional uncertainty on $C_\ell$ agrees with the expected $1/\sqrt{N}$ behavior for $N$ neutrino capture events.  For a full data analysis, not part of this study, one would minimize the number of total bins on the sky map, the total counting statistics for a given highest-$l$ and include angular folding and unfolding studies for neutrino capture events.

The contributions of cosmic variance and counting statistics to the fractional uncertainties on $C_\ell$ are shown in Figure~\ref{fig:Clratio} and compared to the expected effects from WDM.  The rapid growth of the counting statistics uncertainties for high-$l$ has a direct impact on the measurement of WDM effects.  The total number of neutrino captures is nominally limited by the target nuclei capture cross section, \CNB flux and density of states, the total number of nuclei in the target and the acceptance efficiency of the PTOLEMY spectrometer.  However, it may be possible to bring down the counting statistics uncertainties to a level where the WDM effects can be observed in the data, under the linear assumptions made here, with an amplification mechanism, as described in the following section.  Further investigations on the $C_\ell$ predictions including the non-linear matter power spectra are needed.


\section{Quantum Amplification of Neutrino Capture}

The possibility of a superfluid ground-state condensate of spin-polarized tritium atoms may provide an amplification mechanism for the process of relic neutrino capture.  Spin-polarized tritium atoms, which are neutral bosons, have long been considered an ideal candidate to form a quantum liquid~\cite{bevslic2013spin}.  The lightest spin-polarized hydrogen isotope in a gaseous phase was successfully cooled down in a magnetic trap to form a Bose-Einstein Condensate in 1998~\cite{fried1998bose}.  The heavier mass of the tritium reduces the zero-point motion and opens up the possibility of forming a liquid a low temperatures.  Thin films of superfluid spin-polarized tritium adsorbed onto the surface of superfluid $^4He$ may be a suitable target for the PTOLEMY experiment that aims to detect and measure the anisotropies of the \CNB~\cite{marin2013spin,betti2019design}.

\begin{figure}[H]
\begin{center}
\includegraphics[width=0.75\linewidth]{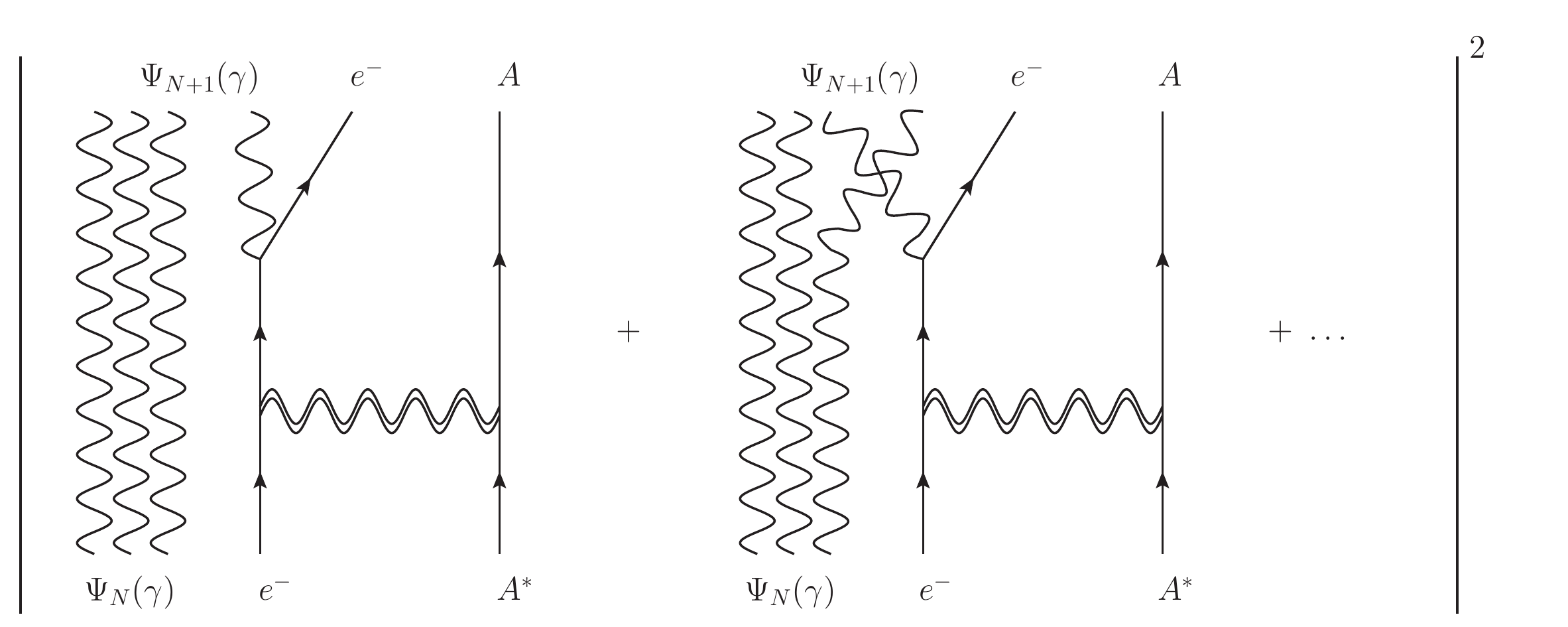}
\includegraphics[width=0.75\linewidth]{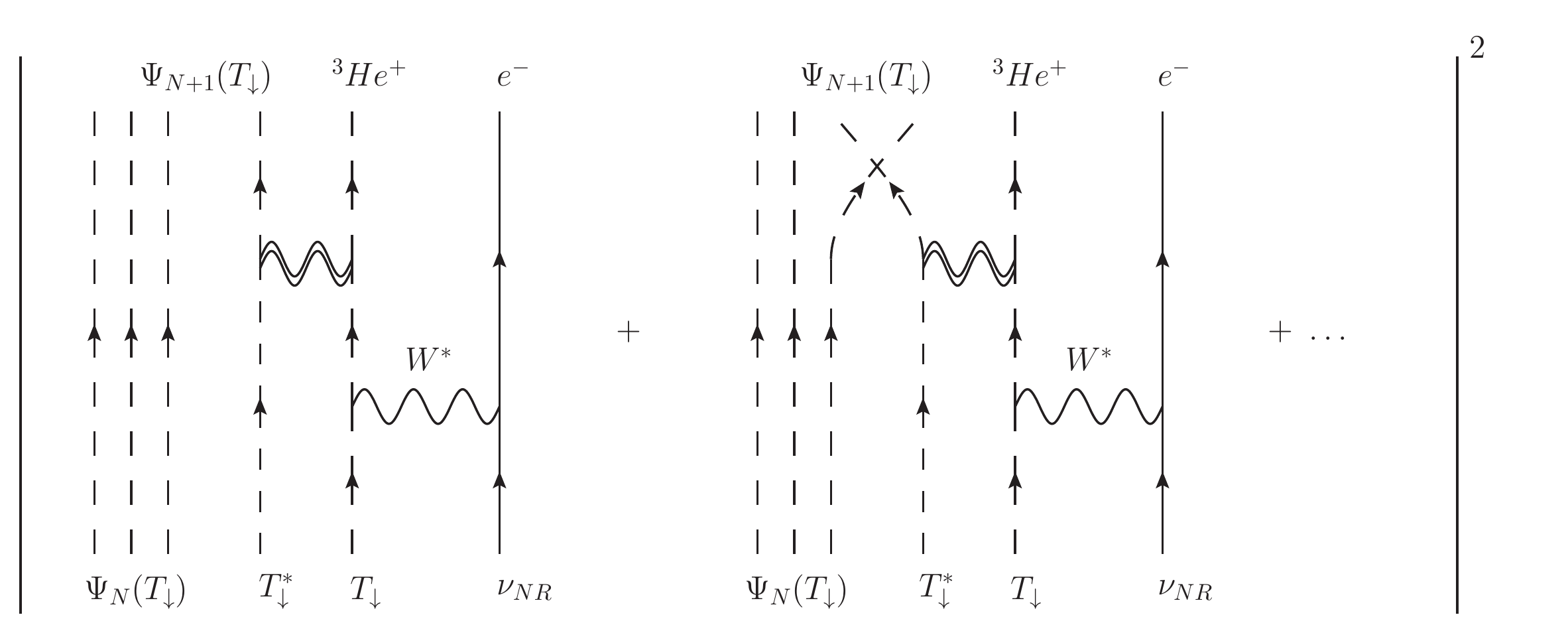}
\caption{Comparison of a laser {\it (top)} to a quantum liquid amplification process for relic neutrino capture {\it (bottom)}.  The amplitudes for exchange terms involving the stimulated increase from $N$ to $N+1$ bosons in the final-state condensate sum coherently with a positive sign following Bose-Einstein statistics.\label{fig:qacnb}}
\end{center}
\end{figure}

The mechanism for quantum liquid amplification of relic neutrino capture is shown diagrammatically in Figure~\ref{fig:qacnb} and compared to the more familiar process of a laser.  In a laser, the long time-constant for the decay of an electron in an excited state of atom $A^*$ is dramatically shortened by the proximity of a photon condensate that shares a final state momentum available to a photon emitted in the de-excitation process of the atomic state.  For a photon emitted into the condensate, the amplitudes for all possible exchange diagrams add with a positive sign for Bose-Einstein statistics and effectively create an amplification of $N+1$ for the de-excitation process in the presence of $N$ indistinguishable photons.  Any process that stimulates the de-excitation or cooling of a non-condensate spin-polarized tritium atom $T_\downarrow^*$ so that it falls into the ground-state $T_\downarrow$ condensate will be amplified in a similar manner.

\unskip

The processes of relic neutrino capture and tritium $\beta$-decay produce an electron and a $^3He^+$ ion in the final state, with the addition of an anti-neutrino for $\beta$-decay.  The $^3He^+$ ion, in particular, has an attractive potential energy with atomic tritium and the ejection of the $^3He^+$ ion from a spin-polarized tritium superfluid has to the potential to provide a cooling mechanism for non-condensate spin-polarized tritium atoms.  A two-fluid model for the tritium superfluid would naturally provide an admixture of ground-state condensate and non-condensate atoms in the liquid phase~\cite{london1939thermodynamics,landau1941theory}.  The coherence length of the superfluid sets the average number $N$ of atoms in a cluster of ground-state condensate.  For each spectator $T_\downarrow^*$ that de-excites into a ground-state cluster, there are $N+1$ coherent amplitudes contributing to the final state.  For symmetric heating/cooling probabilities of the $^3He^+$ ion interacting with the tritium superfluid, one would want more normal than ground-state condensate fraction to have a population inversion.  For the case of relic neutrino capture, the similarity of the process with tritium $\beta$-decay would allow the amplification factor to be determined directly from the shortening of the lifetime of spin-polarized tritium atoms in the superfluid.

The wavelengths of relic neutrinos are macroscopic.
The typical momentum of a relic neutrino averaged over the density of states of a relativistic Fermi-Dirac distribution is roughly $p_\nu = 3 k_B T_\nu/c \approx 0.5$\,meV/$c$.  The corresponding de~Broglie wavelength is $\lambda_\nu = h/p_\nu \approx 3$\,mm.  While the wavelength of the relic neutrino does not directly factor into the cooldown of non-condensate, spectator spin-polarized tritium atoms (coherent neutral current scattering is suppressed by the weak interaction), the probability of a $T_\downarrow^*$ spectator to be within $\lambda_\nu$ is high.

The formation of thin films of superfluid spin-polarized tritium atoms has not been achieved experimentally, though one can consider experimental constraints on these systems for use in the PTOLEMY experiment.  For a liquid density (atoms/volume) of 0.008\,\AA$^{-3}$, the thickness of the films should be limited to roughly 500\,\AA, equivalent to roughly 100 layers of hydrogen.  At this thickness, the Poisson probability for tritium endpoint electrons to inelastically scatter before exiting the thin film target is small~\cite{aseev2000energy,abdurashitov2017electron}.  The recombination rate and the heating effects of tritium $\beta$-decay should have a limited impact on the active target performance with a suitable active target liquid recirculating procedure.  Recycling of the superfluid target is needed to regenerate stable operating conditions.  Similarly with the previous constraint, the vacuum above the thin film should be maintained and pumped down to prevent target materials from entering the spectrometer vacuum regions.  Though there are no estimates for the coherence lengths for a spin-polarized tritium superfluid, the coherence length in a 3D liquid up to the thickness of the film could provide as large as $N=(10^2)^3$ atoms in the condensate clusters increasing the equivalent target mass significantly.
The nominal PTOLEMY \CNB detection rate for 100 grams of tritium is 10 events per year, much fewer than needed to measure even the $l=1$ dipole moment.  A significant gain factor on the neutrino capture rate would open up the possibility for more detailed studies of the \CNB anisotropies.  Figure~\ref{fig:Clratio} shows the importance of reducing the uncertainty from counting statistics on the measurement of high $l$-modes in order to measure the effects of WDM.

The quantum amplification of relic neutrino capture, if this possibility can be realized in physical systems, opens up an exciting possibility for precision measurements of the \CNB anisotropies out to high multipole moments in the angular power spectrum.

\section{Results and Conclusions}

At the same angular scales that one can observe baryonic acoustical oscillations in the CMB, the \CNB anisotropies begin to become sensitive to differences in WDM and CDM cosmologies.
Within the limitations of linear power spectra, deviations of the angular power spectra for \CNB show sensitivity to the downward trend in the expected $C_\ell$ toward high-$l$ for sterile neutrino masses of 2 and 3\,keV.  Heavier masses of 7\,keV are comparable to the CDM.  The cosmic variance uncertainties impose a limitation on the expected fluctuations in $C_\ell$, but it may be possible to reduce these effects for the \CNB using 
the multi-messenger approach described in Ref.~\cite{2021JCAP...06..053T}.  By correlating \CNB anisotropies to luminous tracers that track the matter power spectrum, cosmic variance can be replaced by data-derived predictions for the \CNB anisotropies using optical data.  
Extensive galactic surveys, such as SDSS IV/V, track baryonic matter and can be used to provide a more direct handle on the onset of non-linear power effects at low redshift~\cite{kollmeier2017sdss}.

The quantum liquid amplification process described here for enhancing relic neutrino capture increases the effective tritium target mass for a fixed area and a given time exposure with the possibility to provide sufficient counting statistics needed to observe the impact of WDM on the \CNB anisotropies.  The quantum liquid amplification process, if it can be realized in a practical configuration, may also relevant for limited aperture MAC-E neutrino mass experiments and direct keV-mass sterile neutrino searches that look for kinematic thresholds appearing in the tritium spectrum~\cite{aker2021first,mertens2020characterization}.  Effects of mixing may open up the possibility of direct sterile neutrino capture~\cite{li2011possible}.  The superfluid tritium target as a highly delocalized state of atomic tritium also has the potential to reduce zero-point motion effects on the tritium endpoint electron energy distribution~\cite{cheipesh2021heisenberg,nussinov2021quantum}.

\vspace{6pt} 




\acknowledgments

CGT is supported by the Simons Foundation (\#377485).
A deep gratitude to Hector de Vega and Norma Sanchez for their early acknowledgement in 2013 of the future potential of the PTOLEMY experiment and the discussions and interactions that contributed to this work as part of the Ecole Chalonge-deVega.
\bibliographystyle{JHEP}
\bibliography{cnbwdm.bib}

\providecommand{\href}[2]{#2}\begingroup\raggedright\begin{thebibliography}{10}

\bibitem{kamionkowski1999cosmic}
M.~Kamionkowski and A.~Kosowsky, \emph{{The cosmic microwave background and
  particle physics}}, {\emph{Annual Review of Nuclear and Particle Science}
  {\bfseries 49} (1999) 77}.

\bibitem{2021JCAP...06..053T}
C.G.~{Tully} and G.~{Zhang}, \emph{{Multi-messenger astrophysics with the
  cosmic neutrino background}},
  \href{https://doi.org/10.1088/1475-7516/2021/06/053}{\emph{\jcap} {\bfseries
  6} (2021) 053} [\href{https://arxiv.org/abs/2103.01274}{{\ttfamily
  2103.01274}}].

\bibitem{de2017equation}
H.J.~de~Vega and N.G.~Sanchez, \emph{{Equation of state, universal profiles,
  scaling and macroscopic quantum effects in warm dark matter galaxies}},
  {\emph{The European Physical Journal C} {\bfseries 77} (2017) 1}.

\bibitem{hipolito2018general}
W.S.~Hip{\'o}lito-Ricaldi, R.~vom Marttens, J.~Fabris, I.~Shapiro and
  L.~Casarini, \emph{{On general features of warm dark matter with reduced
  relativistic gas}}, {\emph{The European Physical Journal C} {\bfseries 78}
  (2018) 1}.

\bibitem{class2011}
D.~{Blas}, J.~{Lesgourgues} and T.~{Tram}, \emph{{The Cosmic Linear Anisotropy
  Solving System (CLASS). Part II: Approximation schemes}},
  \href{https://doi.org/10.1088/1475-7516/2011/07/034}{\emph{\jcap} {\bfseries
  2011} (2011) 034} [\href{https://arxiv.org/abs/1104.2933}{{\ttfamily
  1104.2933}}].

\bibitem{dakin2019}
J.~{Dakin}, J.~{Brandbyge}, S.~{Hannestad}, T.~{Haugb{\O}lle} and T.~{Tram},
  \emph{{{\ensuremath{\nu}}CONCEPT: cosmological neutrino simulations from the
  non-linear Boltzmann hierarchy}},
  \href{https://doi.org/10.1088/1475-7516/2019/02/052}{\emph{\jcap} {\bfseries
  2019} (2019) 052} [\href{https://arxiv.org/abs/1712.03944}{{\ttfamily
  1712.03944}}].

\bibitem{viel2012non}
M.~Viel, K.~Markovi{\v{c}}, M.~Baldi and J.~Weller, \emph{{The non-linear
  matter power spectrum in warm dark matter cosmologies}}, {\emph{Monthly
  Notices of the Royal Astronomical Society} {\bfseries 421} (2012) 50}.

\bibitem{peebles1980large}
{Peebles, P.J.E}, \emph{{The large-scale structure of the universe}},
  {\emph{Princeton University Press, 435 p,} (1980) }.

\bibitem{bevslic2013spin}
I.~Be{\v{s}}li{\'c}, L.~Vranje{\v{s}}~Marki{\'c} and J.~Boronat,
  \emph{{Spin-polarized hydrogen and its isotopes: a rich class of quantum
  phases}}, {\emph{Low Temperature Physics} {\bfseries 39} (2013) 857}.

\bibitem{fried1998bose}
D.G.~Fried, T.C.~Killian, L.~Willmann, D.~Landhuis, S.C.~Moss, D.~Kleppner
  et~al., \emph{{Bose-Einstein condensation of atomic hydrogen}},
  {\emph{Physical Review Letters} {\bfseries 81} (1998) 3811}.

\bibitem{marin2013spin}
J.~Mar{\'\i}n, L.V.~Marki{\'c} and J.~Boronat, \emph{{Spin-polarized hydrogen
  adsorbed on the surface of superfluid $^4He$}}, {\emph{The Journal of
  Chemical Physics} {\bfseries 139} (2013) 224708}.

\bibitem{betti2019design}
M.~Betti, M.~Biasotti, A.~Bosc{\'a}, F.~Calle, J.~Carabe-Lopez, G.~Cavoto
  et~al., \emph{A design for an electromagnetic filter for precision energy
  measurements at the tritium endpoint}, {\emph{Progress in Particle and
  Nuclear Physics} {\bfseries 106} (2019) 120}.

\bibitem{london1939thermodynamics}
H.~London, \emph{{Thermodynamics of the thermomechanical effect of liquid He
  II}}, {\emph{Proceedings of the Royal Society of London. Series A.
  Mathematical and Physical Sciences} {\bfseries 171} (1939) 484}.

\bibitem{landau1941theory}
L.~Landau, \emph{{Theory of the superfluidity of helium II}}, {\emph{Physical
  Review} {\bfseries 60} (1941) 356}.

\bibitem{aseev2000energy}
V.~Aseev, A.~Belesev, A.~Berlev, E.~Geraskin, O.~Kazachenko, Y.E.~Kuznetsov
  et~al., \emph{{Energy loss of 18 keV electrons in gaseous T and quench
  condensed D films}}, {\emph{The European Physical Journal D-Atomic,
  Molecular, Optical and Plasma Physics} {\bfseries 10} (2000) 39}.

\bibitem{abdurashitov2017electron}
D.~Abdurashitov, A.~Belesev, V.~Chernov, E.~Geraskin, A.~Golubev, G.~Koroteev
  et~al., \emph{{Electron scattering on hydrogen and deuterium molecules at
  14--25 keV by the ``Troitsk nu-mass'' experiment}}, {\emph{Physics of
  Particles and Nuclei Letters} {\bfseries 14} (2017) 892}.

\bibitem{kollmeier2017sdss}
J.A.~Kollmeier, G.~Zasowski, H.-W.~Rix, M.~Johns, S.F.~Anderson, N.~Drory
  et~al., \emph{{SDSS-V: pioneering panoptic spectroscopy}}, {\emph{arXiv
  preprint arXiv:1711.03234} (2017) }.

\bibitem{aker2021first}
M.~Aker, A.~Beglarian, J.~Behrens, A.~Berlev, U.~Besserer, B.~Bieringer et~al.,
  \emph{{First direct neutrino-mass measurement with sub-eV sensitivity}},
  {\emph{arXiv preprint arXiv:2105.08533} (2021) }.

\bibitem{mertens2020characterization}
S.~Mertens, T.~Brunst, M.~Korzeczek, M.~Lebert, D.~Siegmann, A.~Alborini
  et~al., \emph{{Characterization of silicon drift detectors with electrons for
  the TRISTAN project}}, {\emph{Journal of Physics G: Nuclear and Particle
  Physics} {\bfseries 48} (2020) 015008}.

\bibitem{li2011possible}
{Li, Y.F. and Xing, Z.Z}, \emph{Possible capture of kev sterile neutrino dark
  matter on radioactive $\beta$-decaying nuclei}, {\emph{Physics Letters B}
  {\bfseries 695} (2011) 205}.

\bibitem{cheipesh2021heisenberg}
Y.~Cheipesh, V.~Cheianov and A.~Boyarsky, \emph{{Heisenberg's uncertainty as a
  limiting factor for neutrino mass detection in $\beta$-decay}}, {\emph{arXiv
  preprint arXiv:2101.10069} (2021) }.

\bibitem{nussinov2021quantum}
S.~Nussinov and Z.~Nussinov, \emph{{Quantum Induced Broadening-A Challenge For
  Cosmic Neutrino Background Discovery}}, {\emph{arXiv preprint
  arXiv:2108.03695} (2021) }.

\end{thebibliography}\endgroup


%


\end{document}